\documentclass[sigconf,nonacm]{acmart}

\AtBeginDocument{%
  \providecommand\BibTeX{{%
    Bib\TeX}}}

\usepackage{amssymb,amsfonts}
\usepackage{algorithmic}
\usepackage{graphicx}
\usepackage{textcomp}
\usepackage{xcolor}
\usepackage{subcaption}
\usepackage{listings}
\usepackage{multirow}
\usepackage{tabularx}
\usepackage{url}
\newcolumntype{C}{>{\centering\arraybackslash}X}

\def\BibTeX{{\rm B\kern-.05em{\sc i\kern-.025em b}\kern-.08em
    T\kern-.1667em\lower.7ex\hbox{E}\kern-.125emX}}

\newif\ifdraft{}
\drafttrue{}

\ifdraft{}
  \newcommand{\jhanote}[1]{ {\textcolor{red} { ***shantenu: #1 }}}
  \newcommand{\generalnote}[1]{ {\textcolor{gray} { *note: #1 }}}

\else
  \newcommand{\jhanote}[1]{}
  \newcommand{\generalnote}[1]{}
\fi
    
\begin{document}

\title{ In-Transit Data Transport Strategies for Coupled AI-Simulation Workflow Patterns}

\author{Harikrishna Tummalapalli}
\affiliation{%
  \institution{Argonne National Laboratory}
  \city{Lemont}
  \state{IL}
  \country{USA}
}
\email{htummalapalli@anl.gov}

\author{Riccardo Balin}
\affiliation{%
  \institution{Argonne National Laboratory}
  \city{Lemont}
  \state{IL}
  \country{USA}
}
\email{rbalin@anl.gov}

\author{Christine M. Simpson}
\affiliation{%
  \institution{Argonne National Laboratory}
  \city{Lemont}
  \state{IL}
  \country{USA}
}
\email{csimpson@anl.gov}

\author{Andrew Park}
\affiliation{%
  \institution{Rutgers University}
  \city{New Brunswick}
  \state{NJ}
  \country{USA}
}
\email{andrew.k.park@rutgers.edu}

\author{Aymen Alsaadi}
\affiliation{%
  \institution{Rutgers University}
  \city{New Brunswick}
  \state{NJ}
  \country{USA}
}
\email{aymen.alsaadi@rutgers.edu}

\author{Andrew E. Shao}
\affiliation{%
  \institution{Hewlett Packard Enterprise}
  \city{Victoria}
  \state{BC}
  \country{Canada}
}
\email{andrew.shao@hpe.com}

\author{Wesley Brewer}
\affiliation{%
  \institution{Oak Ridge National Laboratory}
  \city{Oak Ridge}
  \state{TN}
  \country{USA}
}
\email{brewerwh@ornl.gov}

\author{Shantenu Jha}
\affiliation{%
  \institution{Rutgers University}
  \city{New Brunswick}
  \state{NJ}
  \country{USA}
}
\affiliation{%
  \institution{Princeton Plasma Physics Laboratory}
  \city{Princeton}
  \state{NJ}
  \country{USA}
}
\email{shantenujha@acm.org}

\renewcommand{\shortauthors}{Tummalapalli et al.}

\begin{abstract}
Coupled AI-Simulation workflows are becoming the major workloads for HPC facilities, and their increasing complexity necessitates new tools for performance analysis and prototyping of new in-situ workflows. We present SimAI-Bench, a tool designed to both prototype and evaluate these coupled workflows. In this paper, we use SimAI-Bench to benchmark the data transport performance of two common patterns on the Aurora supercomputer: a one-to-one workflow with co-located simulation and AI training instances, and a many-to-one workflow where a single AI model is trained from an ensemble of simulations. For the one-to-one pattern, our analysis shows that node-local and DragonHPC data staging strategies provide excellent performance compared Redis and Lustre file system. For the many-to-one pattern, we find that data transport becomes a dominant bottleneck as the ensemble size grows. Our evaluation reveals that file system is the optimal solution among the tested strategies for the many-to-one pattern.
\end{abstract}




%
%
\maketitle
\section{Introduction}
\label{sec:intro}
%
As the impact of artificial intelligence (AI) deepens in every scientific domain, the nature of high-performance computing (HPC) is undergoing a radical transformation. Monolithic HPC applications are giving way to HPC workflows with many diverse tasks and complex execution patterns. In particular, workflows with a mix of traditional HPC, training and inference tasks\cite{breweraicoupled2024,bard2023workflow,jha2022ai_v2} are becoming standard practice for accelerating scientific campaigns with AI. However, as both simulation and AI models grow to exascale, training AI models from simulation data stored on disk (offline training) is becoming prohibitively expensive. Hence, AI models are increasingly being trained in situ or in transit (generally referred to online) alongside simulations.

Coupled AI-simulation workflows are expected to become an essential part of scientific computing \cite{breweraicoupled2024,bard2023workflow}. Some examples include deep learning for real-time event detection in particle physics experiments \cite{boggia2025review}, steering molecular dynamics simulations \cite{lee2019deepdrivemd}, and building AI surrogates on-the-fly from computational fluid dynamics data \cite{balin2023situ}. Furthermore, it is expected that use of AI agents to drive these online workflows will become an important tool in the near future. 

At the core of AI-coupled HPC workflows is a need for data transport between components and a high-degree of data reuse, as highlighted by recent efforts to categorize coupled workflows into common motifs \cite{breweraicoupled2024}. As a consequence, data streaming and staging solutions, which can avoid costly I/O to the parallel file system, such as ADIOS2 \cite{godoy2020adios}, DataSpaces \cite{docan2010dataspaces}, SmartSim \cite{partee2022,balin_gnn}, and DragonHPC \cite{DragonHPC}, have emerged as common tools for obtaining performance at scale. Nevertheless, data transport and staging can still be a major bottleneck for workflow performance \cite{suter2025terminology,breweraicoupled2024}. For example, building a surrogate model from an ensemble of computational fluid dynamics (CFD) simulations requires significant data movement from the simulation instances to the AI training process, running into bandwidth limitations of the system interconnect. Alternatively, inference workloads can be latency limited, with the cost of data transfer dominating over the computational one. 

Despite the growing adoption of coupled simulation-AI workflows and the need for performant data transport, scientists still face a number of implementation and deployment challenges when developing new applications or optimizing existing ones. From the highest level of workflow management to the specific streaming or staging approach to be used, the lack of clear guidelines or best practices coupled with the shear number of tools available (over 300 \cite{workflow_list}) creates a significant barrier to entry. Additionally, due to the non-trivial effort needed to implement such workflows, performance and portability studies comparing various tools with the same execution motifs across multiple systems are lacking.
 
In HPC, computational scientists have often used mini-apps-small, self-contained programs that isolate key computational aspects without the complexity of the full application-in order to benchmark hardware and software. In the context of coupled AI-simulation workflows, a similar argument can be used to derive mini-apps for end-to-end workflows capable of modeling the key task execution and data transfer patterns. However, generating representative mini-apps for full coupled workflows is significantly more complicated than for traditional, monolithic HPC applications. This difficulty arises because a workflow mini-app must emulate not only the computational components (e.g., solvers or ML models), but also the complex interaction patterns between them. For example, such workflows may involve one-to-one or many-to-many communication, be static or dynamic in nature, or require intricate task and data dependencies \cite{breweraicoupled2024}. Capturing such a wide spectrum of possibilities is a software challenge. 
Moreover, traditional HPC applications rely heavily on distributed synchronous computations to achieve scientific outcomes, whereas AI-integrated workflows rely on concurrent but asynchronous and highly varied computations, often running on different hardware. Workflows, therefore, abandon the common single program, multiple data (SPMD) programming model in favor of a multiple program, multiple data approach with new paradigms for distributed computing. 


To address these gaps, we introduce SimAI-Bench \cite{simaibench} as a flexible and extensible interface for emulating, benchmarking, and prototyping AI-coupled HPC workflows, including data transport. By providing a set of Python packages and intuitive API, SimAI-Bench allows users to compose faithful mini-apps of full end-to-end workflows with diverse simulation and AI components and multiple in situ or in transit data transfer patterns. In this work we use SimAI-Bench to develop mini-apps of workflows performing online training of an AI surrogate from a single parallel simulation or an ensemble of simulations. This pattern was chosen because it appears in many of the workflow motifs identified in \cite{breweraicoupled2024}. In particular, these mini-apps allow us to benchmark the performance of different deployment strategies and implementations for data transport up to 512 nodes of the Aurora supercomputer at the Argonne Leadership Computing Facility. 
With this work, we make the following contributions:
\begin{enumerate}
    \item We introduce SimAI-Bench, describe its various components and how they can be used to generate workflow mini-apps. 
    \item Through comparisons to a real HPC production workflow, we demonstrate how SimAI-Bench mini-apps can faithfully represent the data transfer patterns and interactions between components.
    \item With the flexible design of SimAI-Bench, we demonstrate how mini-apps can provide insightful performance information for various data transport tools and patterns.
\end{enumerate}

The paper is organized as follows. Section~\ref{sec:related_work} discusses related work in this field in order to ground our contributions. Section~\ref{sec:Design} covers the design of SimAI-Bench and its various components. Results obtained with our workflow mini-apps are presented in Section~\ref{sec:evaluation} and final conclusions and opportunities for future work are included in Section~\ref{sec:conclusion}.
\section{Related work}
\label{sec:related_work}
%
The use of mini-apps for performance analysis and benchmarking is a well-established practice in HPC \cite{crozier_improving_2009}. This practice is motivated by the observation that in many large scientific applications, a few computational kernels account for the majority of the runtime. Mini-apps are typically designed to match the floating-point operations per second (FLOPS) and memory access patterns of the production codes they represent \cite{messer_miniapps_2018}, serving a wide variety of purposes, including optimizing application performance \cite{crozier_improving_2009}, co-designing system architectures \cite{saurabh_quantum_2024}, benchmarking new systems \cite{barrett_assessing_2015}, and evaluating programming models \cite{kwack_evaluation_2021}.

Perhaps the most widely known HPC benchmark is the Top500 HPL benchmark \cite{luszczek2006hpc}. Benchmarking suites more representative of full production codes also exist; for example, the Mantevo project \cite{crozier_improving_2009} includes applications for finite element analysis (MiniFE), molecular dynamics (MiniMD), and circuit simulation (MiniXyce). The Mantevo project also includes a configurable proxy application to emulate the performance characteristics of other applications. Similar domain-specific mini-apps have been developed for physics \cite{martineau_arch_2017}, scientific data analysis \cite{sukumar_mini-apps_2016}, and ab-initio molecular dynamics \cite{mas_magre_nomad_2025}. Recently, the scope of mini-apps has expanded to address modern challenges like the performance portability of programming models across different architectures \cite{kwack_evaluation_2021}. Relevant to our work, \cite{marts_minimod_2021} designed a framework to execute traditional mini-app kernels under various communication and threading models, highlighting a trend toward more flexible and configurable benchmarking tools.

Given the increasing role of workflows in scientific workloads, efforts have recently been made to create representative workflow mini-apps (or simply mini-apps in this context). These are intended to model the behavior of full workflows, not just the individual components. For instance, WfBench \cite{coleman2022wfbench} is a tool designed to automate the generation of workflow benchmarks that emulate real applications in terms of CPU, memory, and I/O usage. This package uses WfChef \cite{coleman2021wfchef} to generate realistic task-dependency graphs and configures the individual tasks to match the performance characteristics of a target application. However, WfBench has two main limitations relevant to our work. First, data is exchanged between tasks exclusively through files on disk; in transit data transfer through available memory or through the interconnect is not considered. Second, it currently only supports the emulation of CPU-based tasks. Another approach described in \cite{kilic_workflow_2024} introduces workflow mini-apps built using a library of specific, tunable kernels. This library provides kernels for both CPU and GPU that can be configured to match performance metrics like makespan and resource utilization of a real workflow. Notably, this approach focuses only on emulating the tasks themselves and explicitly relies on external workflow management systems, such as RADICAL-Cybertools, for task orchestration. As such, it does not provide an integrated task orchestration or a specific data transport layer. Additionally, it is not integrated with ML framework libraries such a PyTorch, thus emulating ML components through basic GEMM kernels.

Coupled AI-Simulation interaction patterns involve a significant amount of data transport between components, which warrants high-performance data exchange mechanisms. The scientific workflow community has developed a detailed terminology to categorize these mechanisms, distinguishing between different data transport methods (file-based, streaming, in-memory) and storage architectures (shared, distributed, replicated) \cite{suter2025terminology}. While many existing workflow systems still rely on traditional file-based transport for its simplicity and robustness, there is a clear trend toward more sophisticated data management and streaming approaches in response to growing data volumes and the demand for low-latency, online processing \cite{suter2025terminology}. 

Considering this evolution, it is imperative that workflow benchmarks and mini-apps also evolve to represent both critical control and data plane characteristics of full end-to-end workflows, unlike the traditional HPC mini-apps. Moreover, providing support for both traditional HPC kernels and AI/ML framework is essential for benchmarking simulation-AI coupled workflows. This is the specific gap SimAI-Bench is designed to fill by providing a framework to prototype and benchmark full end-to-end workflows using portable mini-apps.

\section{Design}
\label{sec:Design}
\begin{figure}[htbp]
    \centering
    \includegraphics[width=\linewidth]{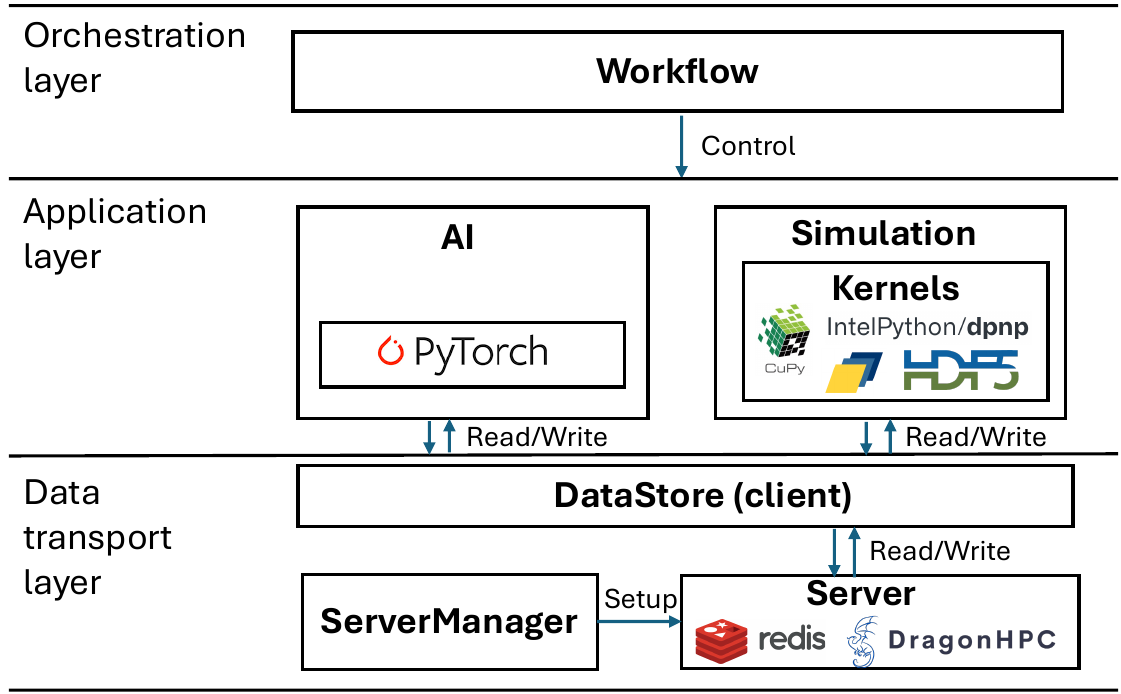}
    \caption{Architecture of SimAI-Bench.}
    \label{fig:fig1_architecture}
    \Description{
This image displays the three-tiered software architecture of SimAI-Bench for integrated AI and Simulation workflows. The top Orchestration layer contains a Workflow component that directs the system via a "Control" flow to the layer below. This middle Application layer is divided into an AI block, which uses PyTorch, and a Simulation block containing Kernels built with CuPy, IntelPython/dpnp, and HDF5. Both application blocks interact with the bottom Data Transport layer via "Read/Write" operations. This final layer manages data movement using a DataStore and a ServerManager, which operate on top of a foundational Server composed of Lustre, Redis, and DragonHPC. The DataStore has a "Read/Write" connection to the Server, while the ServerManager performs a "Setup" function on it.
    }
\end{figure}

This section details the architecture of SimAI-Bench, a modular framework designed to construct and benchmark workflow mini-apps.   A schematic of the component stack is shown in Fig. \ref{fig:fig1_architecture}.
At the highest level, the \texttt{Workflow} class provides the orchestration layer, enabling users to define workflows by setting dependencies between components. The primary actors within a workflow are the \texttt{Simulation} and \texttt{AI} classes. The \texttt{Simulation} class emulates traditional scientific solvers by using primitives from the \texttt{Kernels} module, while the \texttt{AI} class uses \texttt{PyTorch} models to emulate ML training and inference tasks. Both components rely on the \texttt{DataStore} module to provide a unified interface for staging and streaming data.

\subsection{Kernels}
The \texttt{Kernels} module provides primitives for key operations, including compute, I/O, collective communication, and copy. It is adapted from \texttt{workflow-mini-api}~\cite{kilic_workflow_2024} and has been extended to support Intel's \texttt{dpnp} \cite{dpnp} library for efficient execution on Aurora's Intel Data Center Max 1550 GPUs \cite{pvc_paper}. Each type of operation is managed by a specialized library. For compute and copy operations, the kernels leverage \texttt{CuPy} and \texttt{dpnp} to handle execution and device-host data transfers. Collective communications are implemented using \texttt{mpi4py} for MPI and NCCL for GPU-direct communication on NVIDIA GPUs (\texttt{dpnp} does not natively support GPU-GPU communication). Lastly, I/O operations utilize HDF5 for I/O. The module is designed for extensibility, allowing for custom kernels to be easily added. A complete list of the default primitives can be found in Table~\ref{tab:kernel_list}.

\begin{table}[htbp]
\caption{List of kernels provided by the \texttt{Kernel} module.}
\label{tab:kernel_list}
\renewcommand{\arraystretch}{1.4} 
\begin{tabular}{@{} l l >{\raggedright\arraybackslash}p{3.5cm} @{}}
\toprule
\textbf{Category} & \textbf{Kernel} & \textbf{Description} \\
\midrule
\multirow{7}{*}{Compute}
    & \texttt{MatMulSimple2D}    & Simple 2D matrix multiplication \\
    & \texttt{MatMulGeneral}     & General matrix multiplication (GEMM) \\
    & \texttt{FFT}               & Fast Fourier Transform \\
    & \texttt{AXPY}              & Scalar-vector multiplication and addition ($ax+y$) \\
    & \texttt{InplaceCompute}    & Performs a computation on data in-place ($f(x)$) \\
    & \texttt{GenerateRandomNumber} & Generates an array of random numbers \\
    & \texttt{ScatterAdd}        & Scatters and adds values to an array \\
\cmidrule(lr){1-3}
\multirow{5}{*}{IO}
    & \texttt{WriteSingleRank}   & A single process writes data to a file \\
    & \texttt{WriteNonMPI}       & Writes data to a file without MPI-IO \\
    & \texttt{WriteWithMPI}      & Writes data using MPI-IO collectives \\
    & \texttt{ReadNonMPI}        & Reads data from a file without MPI-IO \\
    & \texttt{ReadWithMPI}       & Reads data using MPI-IO collectives \\
\cmidrule(lr){1-3}
\multirow{2}{*}{Collectives}
    & \texttt{AllReduce}         & Performs an all-reduce operation \\
    & \texttt{AllGather}         & Performs an all-gather operation \\
\cmidrule(lr){1-3}
\multirow{2}{*}{Copy}
    & \texttt{CopyHostToDevice}  & Copies data from CPU to GPU memory \\
    & \texttt{CopyDeviceToHost}  & Copies data from GPU to CPU memory \\
\bottomrule
\end{tabular}
\end{table}

\subsection{ServerManager and DataStore}
\label{subsec:datastore}
The \texttt{ServerManager} and \texttt{DataStore} classes are designed to collectively provide a unified interface for data staging and streaming through various approaches and software libraries, heretofore referred to as data transport backends. The \texttt{ServerManager} is responsible for the creation and configuration of data servers, while the \texttt{DataStore} exposes a uniform client Application Programming Interface (API) for data transport.

The framework supports four distinct backend types. These include Redis, a popular high-performance in-memory key-value store used extensively in production environments and leveraged by the SmartSim library; DragonHPC, a HPC-specific runtime developed by HPE which features a distributed, in-memory data dictionary \cite{DragonHPC}; high-performance node-local flash storage or RAM memory, for instance SSD or RAM based tmpfs on each compute node; and a parallel file system, such as Lustre.

The \texttt{ServerManager} setup procedure is dependent upon the selected backend. For in-memory stores such as Redis and DragonHPC, it can deploy servers on a given set of compute nodes either as distinct instances or as a cluster. Conversely, for the node-local and file system backends, its function is to establish the necessary directory structures for data staging.

The \texttt{DataStore} class presents a unified client API for all backends, comprising four primary functions: \texttt{stage\_read}, \texttt{stage\_write}, \texttt{poll\_staged\_data}, and \texttt{clean\_staged\_data}. In the case of the node-local and file system backends, a simple sharded key-value store is implemented. Keys are hashed using the CRC32 algorithm to determine a shard directory. To ensure atomicity and prevent race conditions, the value is first written to a temporary file, which is then atomically renamed to its final destination (\texttt{key.pickle}) using \texttt{os.replace()}. It is worth noting that by providing a single set of client APIs, SimAI-Bench allows benchmarking of the different data transport backends and quick prototyping of mini-apps simply by selecting the appropriate arguments at runtime.


\subsection{Simulation}

The \texttt{Simulation} class is designed to emulate the simulation component of a coupled simulation-AI workflow, providing tight integration with the data transport layer to accurately model key performance metrics like makespan and resource placement.

A simulation is defined by a configuration, expressed as a Python dictionary or JSON file as in the example in Listing~\ref{lst:sim_config2}, that abstracts the workflow into a series of constituent kernels. This configuration specifies the essential execution parameters for each kernel, such as the \texttt{mini\_app\_kernel}, which defines the name of the kernel to use, the target device (e.g., CPU or GPU) to model resource placement, and the \texttt{data\_size} involved. The computational workload of each kernel can be characterized deterministically through a fixed \texttt{run\_time} or a specific \texttt{run\_count} of iterations.

To better emulate workloads with variable performance, SimAI-Bench allows the parameters—\texttt{run}\texttt{\_time} and \texttt{run}\texttt{\_count}—to be defined stochastically. Users can provide a discrete probability density function (PDF) for these parameters. At each iteration of the simulation, the execution engine samples from the given distribution, enabling a more realistic emulation of dynamic and unpredictable system behaviors. This configured sequence of operations is then executed using the underlying \texttt{Kernels} module for computation emulation and the \texttt{DataStore} module to manage interactions with the data staging backends.

\subsection{AI}
\label{subsec:AI}
In addition to emulating traditional simulation components, our framework includes a dedicated \texttt{AI} class to model the machine learning portions of a workflow by leveraging the \texttt{torch.nn} and \texttt{torch.distributed} modules in PyTorch. The rationale for a separate class is to encapsulate and emulate the complex compute and communication characteristics inherent to deep learning frameworks, which are often treated as a black box given the extensive number of high-level APIs present in ML frameworks (e.g., \texttt{loss.backwards()} in PyTorch DDP hides both computation and communication to the user). Similar to the \texttt{Simulation} class, the \texttt{AI} class can interact with the \texttt{DataStore} backend for data staging. Furthermore, it offers analogous execution control, allowing training or inference to proceed for a prescribed number of iterations or a specific duration, which mirrors the \texttt{run\_count} and \texttt{run\_time} parameters used by the simulation kernels. The current implementation supports distributed data-parallel training with DDP from \texttt{torch.distributed}, with an initial focus on a feed-forward, fully-connected neural network models. Future work will expand these capabilities to include more advanced architectures, such as graph or convolutional neural networks which are commonly adopted by the scientific ML community.

\subsection{Workflow}
To orchestrate the execution of coupled simulation and AI tasks, our framework introduces a high-level \texttt{Workflow} abstraction. This abstraction is designed around three core architectural principles: modular components, an explicit dependency graph, and an explicit data staging interface. This architectural separation is intentional, as it decouples the logical structure of the workflow from the physical data transport mechanisms. This decoupling enhances both modularity and the flexibility to swap data staging backends without altering the execution of the workflow.

Listing \ref{lst:workflow} illustrates these principles. The \texttt{@w.component} decorator registers a function as a component and specifies critical execution metadata. For example, the type argument directs the workflow manager to deploy the component on remote compute nodes via \texttt{mpirun} ('remote') or on the local node using \texttt{multiprocessing} ('local'). Inter-component relationships are defined explicitly using the \texttt{dependencies} argument, which allows the framework to automatically construct a Directed Acyclic Graph (DAG) of tasks. In the example, this ensures \texttt{run\_sim} is completed before \texttt{run\_sim2} begins.

Data flow between distributed components is handled explicitly via \texttt{stage\_write} and \texttt{stage\_read} calls. This decouples the computational logic from the underlying data transport layer, which is enabled by the \texttt{DataStore} client class. Whereas, the \texttt{ServerManager} that provides a unified interface to setup various data staging backends. Finally, the \texttt{w.launch()} method serves as the entry point to initiate the workflow execution. Once invoked, control is passed to an underlying execution manager that traverses the constructed DAG, dispatches components to their designated resources, and manages their lifecycle until the entire workflow is complete.

While the built-in \texttt{Workflow} class is intended for rapid prototyping, our framework's modular design allows for components developed with the \texttt{Simulation} and \texttt{AI} modules to be exported for use with third-party workflow managers, such as RADICAL-Pilot or Parsl, enabling users to leverage more advanced functionalities.

\begin{lstlisting}[float, language=Python, label={lst:workflow}, xleftmargin=0.5cm, caption={Example of creating and running a workflow mini-app with SimAI-Bench.}]
from SimAIBench import Workflow
from SimAIBench import Simulation
from SimAIBench import ServerManager

server = ServerManager("server", config=server_config)
server.start_server()
info=server.get_server_info()

w = Workflow(sys_info=sys_config)

@w.component(name="sim",type="remote",
             args={"info":info})
def run_sim(info=None):
  sim = Simulation(name="sim",
             server_info=info)
  sim.add_kernel("MatMulSimple2D")
  sim.run()
  sim.stage_write("key1","value1")
  value = sim.stage_read("key2")

@w.component(name="sim2", type="local",
                 args={"info":info},
                 dependencies=["sim"])
def run_sim2(info=None):
  sim = Simulation(name="sim2",
                        server_info=info)
  sim.add_kernel("MatMulGeneral")
  value = sim.stage_read("key1")
  sim.stage_write("key2","value2")
  sim.run()

w.launch()

server.stop_server()
\end{lstlisting}

\section{Evaluation}
\label{sec:evaluation}

The workflow mini-apps evaluated in this section were constructed using the SimAI-Bench framework as detailed in Section~\ref{sec:Design}. Our evaluation is designed to demonstrate how SimAI-Bench can be used to benchmark and optimize data transport for two common patterns observed in online training workflows: a one-to-one pattern were a single parallel simulation trains a single distributed model, and a many-to-one pattern where many concurrent simulations train a single model. All the below experiments were conducted on Aurora supercomputer. 

Aurora is 10,624-node HPE Cray-Ex based system consisting of 166 racks with 21,248 CPUs and 63,744 GPUs. Each node consists of 2 Intel Xeon CPU Max Series and 6 Intel Data Center GPU Max Series. Each Xeon CPU has 52 physical cores supporting 2 hardware threads per core, and each datacenter GPU is further divided into 2 tiles or stacks. Each CPU socket has 512 GB of DDR5 memory and 64 GB of HBM memory.

\subsection{Pattern 1: One-to-One}
\label{subsec:evals_pattern1}

As our first use case, we use SimAI-Bench to model the nekRS-ML workflow, which couples the high-order spectral element CFD solver nekRS \cite{fischer2022nekrs} with a Graph Neural Network (GNN) surrogate model \cite{balin_gnn}. In this workflow pattern, the nekRS simulation periodically produces snapshots of the flow field to be used for online GNN training with the goal of performing efficient forecasting of the flow solution with the surrogate. 
The two components are co-located on the same Aurora nodes, with the 12 tiles of the Data Center Max 1550 GPUs on each node split evenly between the simulation (6 tiles) and the AI trainer (6 tiles).

During the run, the nekRS and GNN components execute concurrently. The data transport pattern is fully asynchronous, meaning that nekRS and the trainer send and receive data completely independently of each other. As nekRS produces training data at periodic intervals (100 iterations), the GNN trainer reads new data at a regular interval of 10 iterations to update its data loader. This process continues until the GNN completes a set number of training iterations (5000 in our tests), at which point it steers the workflow by instructing the nekRS component to stop.

The original production workflow is implemented using the SmartSim library \cite{partee2022}, which relies on a Redis database for data transport and staging. However, it is unclear if this is the most optimal strategy, or how the choice of an optimal backend might change as the data intensity increases. For example, the simulation problem size could grow (i.e., more grid points per rank), or the number of features could increase as the AI model is trained on more fields (e.g., both velocity and pressure, instead of only velocity). To investigate these questions, the following sections use SimAI-Bench to benchmark the performance of multiple data transport backends across a wide range of message sizes.

\subsubsection{Validation}
\label{subsubsec:validation}

To construct the mini-app for this workflow, our primary goal was to faithfully reproduce the makespan, resource placement, and data transport characteristics of the original application, rather than the scientific computation itself. We first profiled a production run of the workflow—consisting of ~10,000 solver iterations and 5000 GNN training iterations—to determine the average iteration and initialization times. These times were measured by placing timers at the start and end of each iteration and computing the mean.

Based on these measurements, the Simulation component of our mini-app was configured to emulate the nekRS iteration time using the \texttt{MatMulSimple2D} kernel. In addition, the Simulation component uses \texttt{run\_time}, \texttt{data\_size}, and \texttt{device} parameters to determine the kernel iteration time, array dimensions for the matrix multiplication, and resource placement, respectively. Listing \ref{lst:sim_config2} presents the configuration used for the current experiment. Similarly, the AI component uses a lightweight feed-forward neural network to match the GNN training time of 0.061 seconds per iteration. This strategy ensures that the emulated components execute for the same duration and on the same GPU tiles as the original application, providing a high-fidelity benchmark of the data transport performance.

We acknowledge our simplified kernels do not capture the full computational complexity of scientific applications. However, as our work focuses on data transport, they are designed to accurately emulate the iteration timings and data volumes that create a realistic I/O load. Modeling other application behaviors, such as arithmetic intensity and portability across different hardware, remains future work.

\begin{lstlisting}[float, language=Python, label={lst:sim_config2}, xleftmargin=0.5cm, caption={Configuration of the simulation component emulating nekRS.}]
{
    "kernels":[
        {
            "name":"nekrs_iter",
            "run_time":0.03147,
            "data_size":[256,256],
            "mini_app_kernel":"MatMulSimple2D",
            "device":"xpu"
        }
    ]
}
\end{lstlisting}

To validate the fidelity of our mini-app, we compared it against the original nekRS-ML workflow using three metrics: global event counts, mean iteration time, and a direct visual timeline.

First, to provide a global statistic of the mini-app's fidelity, Table \ref{tab:number_of_events} presents a comparison of the number of time steps and data transport events between the original workflow and the mini-app. There is an excellent match between the real workflow and the mini-app in terms of the number of events. The small differences are due to the asynchronous nature of the data transport, where the AI component periodically polls for new data.

\begin{table}[htbp]
\renewcommand{\arraystretch}{1.2}
\begin{tabularx}{\columnwidth}{@{} l C C C C @{}}
\toprule
& \multicolumn{2}{c}{\textbf{Simulation}} & \multicolumn{2}{c}{\textbf{Training}} \\
\cmidrule(lr){2-3} \cmidrule(lr){4-5}
& timestep & data transport & timestep & data transport \\
\midrule
Original & 10108 & 203 & 5000 & 208    \\
Mini-app & 10507 & 211 & 5000 & 208 \\
\bottomrule
\end{tabularx}
\caption{Comparison of number of time steps and data transport events in the original workflow and the mini-app. 
}
\label{tab:number_of_events}
\end{table}

Second, Table \ref{tab:mean_std_shooting_wf_tabularx} shows the comparison of the mean and standard deviation of the iteration times obtained from the real workflow and the mini-app. Once again, the mini-app shows excellent agreement with the real workflow statistics in terms of the mean. Comparing the standard deviation of mini-app with the original workflow, it is clear that the mini-app strictly maintains the iteration time close to the provided value in the configuration. However, iteration times of the real workflow  have significant standard deviation, suggesting differences in the probability distributions of the real and mini-app. Despite this, overall data transport read and write events, which are asynchronous and the focus of this study, did not show any significant differences. Consequently, we did not attempt to match the distributions more closely.

\begin{table}[htbp]
\renewcommand{\arraystretch}{1.2}
\begin{tabularx}{\columnwidth}{@{} l C C C C @{}}
\toprule
& \multicolumn{2}{c}{\textbf{Simulation}} & \multicolumn{2}{c}{\textbf{Training}} \\
\cmidrule(lr){2-3} \cmidrule(lr){4-5}
& mean (s) & std (s) & mean (s) & std (s) \\
\midrule
Original & 0.0312 & 0.0273 & 0.0611 & 0.1    \\
Mini-app & 0.0325 & 0.0011 & 0.0633 & 0.0017 \\
\bottomrule
\end{tabularx}
\caption{Comparison of mean and standard deviation (std) of iteration times of the original workflow and the mini-app. 
}
\label{tab:mean_std_shooting_wf_tabularx}
\end{table}

\begin{figure}[htbp]
    \centering
    \begin{subfigure}[b]{\linewidth}
        \centering
        \includegraphics[width=\linewidth]{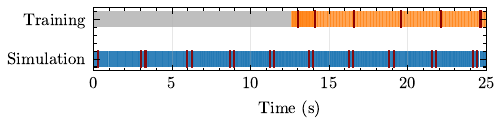}
        \caption{Original}
    \end{subfigure}
    
    \vspace{1em}  

    \begin{subfigure}[b]{\linewidth}
        \centering
        \includegraphics[width=\linewidth]{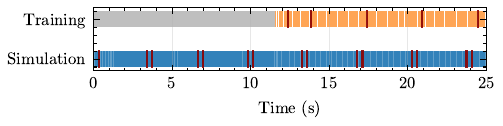}
        \caption{Mini-app}
    \end{subfigure}

    \caption{Comparison of execution timelines for the original nekRS-ML workflow and its SimAI-Bench mini-app replica. Blue and orange regions represent computation for the simulation and training components, respectively. Red bars indicate data transfers (writes by the simulation and reads by the trainer), while gray areas denote initialization time.}
    \Description{This figure displays two timeline charts, labeled (a) Original and (b) Mini-app. Both charts plot activity over a 25-second time interval shown on the horizontal axis. Each chart contains two horizontal tracks: a top track for a 'Training' process and a bottom track for a 'Simulation' process. In the 'Training' track of both charts, an initial gray bar is followed by an orange bar. In the 'Simulation' track, a blue bar spans the full duration. Dark red vertical bars are shown at various points in time within both the Training and Simulation tracks of each chart.
    }
    \label{fig:fig2_timeline}
\end{figure}

To provide a visual validation, Fig. \ref{fig:fig2_timeline} shows a comparison of timelines between the original workflow and the mini-app, showing a segment of the full execution. The timelines clearly show that the mini-app emulates the workflow timeline reasonably well. It successfully reproduces the non-uniformly spaced data transfers that result from the asynchronous pattern mentioned earlier. Considering the significant standard deviation of the original iteration times presented in Table \ref{tab:mean_std_shooting_wf_tabularx}, this close agreement is remarkable.

Overall, the statistics presented in this section provide a comprehensive validation of the mini-app's fidelity. The results confirm that the SimAI-Bench mini-app accurately reproduces not only the high-level characteristics of the workflow, such as the makespan and data transfer frequency, but also the fine-grained statistical behavior of data transport. This ability to faithfully emulate the complex, asynchronous patterns of the original application ensures that the optimization studies that follow are based on a realistic and representative benchmark.

\subsubsection{Optimization}

\begin{figure}[htbp]
    \centering
    \begin{subfigure}[b]{\linewidth}
        \centering
        \includegraphics[width=\linewidth]{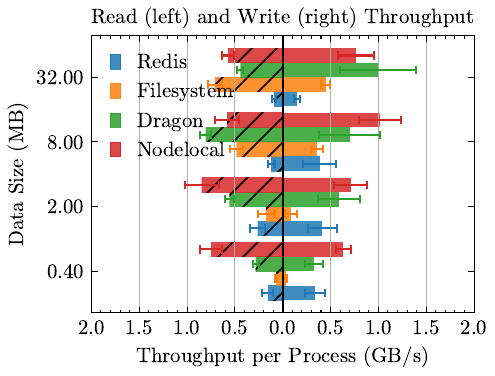}
        \caption{8 nodes}
    \end{subfigure}
    
    \vspace{1em}  

    \begin{subfigure}[b]{\linewidth}
        \centering
        \includegraphics[width=\linewidth]{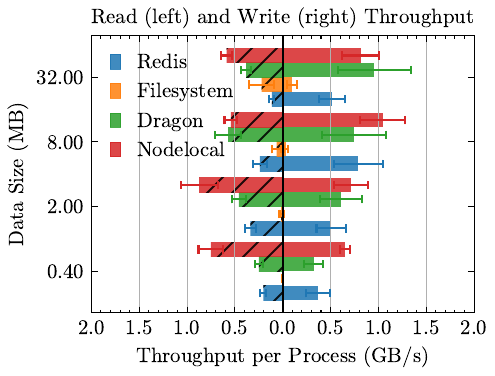}
        \caption{512 nodes}
    \end{subfigure}

    \caption{Read and write throughput with various data transport backends as a function of array size, measured at (a) 8 nodes and (b) 512 nodes of Aurora for Pattern 1. 
    }

    \Description{This figure contains two bar charts, labeled (a) 8 nodes and (b) 512 nodes, that compare the I/O throughput of four different mechanisms. Both charts plot 'Throughput per Process' in GB/s on the horizontal axis against 'Data Size' in MB on the vertical axis. The horizontal axis is centered at zero, with read throughput values extending to the left and write throughput values extending to the right. Each chart displays data for four mechanisms—Redis, Filesystem, Dragon, and Nodelocal—at four different data sizes ranging from 0.40 to 32.00 MB. A set of four horizontal bars is shown for each data size, with the length of each bar corresponding to the measured read and write throughput for a specific mechanism. Error bars are included at the end of each bar.}
    
    \label{fig:fig3_nekrs_ml_rw_throughput}
\end{figure}

To optimize the existing workflow data transport pattern, we investigated the effect of changing the data transport backend. As explained above, the original workflow was implemented using Redis as the data transport backend, where the data is exchanged locally on the node (i.e., simulation ranks only share data with training ranks on the same node). In this test, we compared  Redis with node-local (leveraging the tempfs DRAM based storage on the Aurora nodes), parallel Lustre file system (file system), and DragonHPC distributed dictionary (dragon) solutions. Additionally, we varied the data size starting from 0.4 MB to 32 MB, achieved by increasing the size of arrays containing training data. Note that the original workflow writes/read 1.2 MB of data per rank for every write/read operation. No attempt has been made to optimize the configuration of each of the backends, and used default values. For the Lustre file system, we use a stripe size of 1 MB and count of 1. Each experiment is conducted for at least 2500 training iterations, and rest of the configuration is identical to the validation section. All statistics are obtained by averaging over all the processes and events in the experiment. Fig. \ref{fig:fig3_nekrs_ml_rw_throughput} shows the read and write throughput per process at 8 and 512 nodes on Aurora.

At the 8-node scale, the data transfer throughput is significantly affected by the data size, with different backends showing distinct performance profiles. The in-memory data stores, DragonHPC, Redis, and node-local, exhibit similar non-monotonic behavior. Their throughput for both read and write operations tends to increase with message size before reducing at the largest sizes. Although not presented here, the dragon throughput decreases for data size larger than 32 MB. This performance dip may be attributed to cache effects; the total L3 cache on an Aurora CPU is 105 MB, which provides approximately 8 MB per process in our 12-process per node (6 Simulation + 6 AI) configuration. The largest data transfers exceed this per-process cache, potentially leading to performance-degrading cache spills. In contrast, the parallel file system backend shows a monotonic trend, increasing I/O throughput with data size.

As we scale up to 512 nodes, the performance of the file system degrades significantly, whereas the performance of other backends stays similar. Note that for the file-backed solutions (file system and node-local), the number of shards is increased linearly with the number of nodes. For the Redis, node-local, and DragonHPC solutions, this stable performance is an expected result, as they exchange data locally between the co-located AI and simulation components. For the file system, the significant degradation in performance might be due to metadata overhead from the concurrent requests of 512 nodes creating contention.

Overall, at smaller scales and larger data sizes ($>$8 MB), the file system can be a reasonable choice as a data transport backend, especially considering the additional technical overhead of implementing a staging solution with Redis or DragonHPC which require intrusive changes in the code. In the other scenarios, particularly for large scale runs, either node-local or the DragonHPC backends are the better solution. Redis, even though is an extremely robust solution, is not as performant in terms of throughput.

\begin{figure}[htbp]
    \centering

    \begin{subfigure}[b]{0.48\linewidth}
        \centering
        \includegraphics[width=\linewidth]{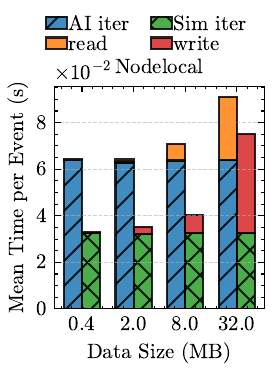}
        \caption{8 nodes}
    \end{subfigure}
    \begin{subfigure}[b]{0.48\linewidth}
        \centering
        \includegraphics[width=\linewidth]{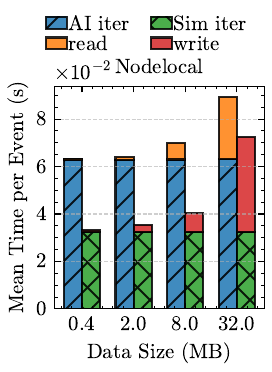} 
        \caption{512 nodes}
        \label{fig:top_right}
    \end{subfigure}

    \vspace{1em} 

    \begin{subfigure}[b]{0.48\linewidth}
        \centering
        \includegraphics[width=\linewidth]{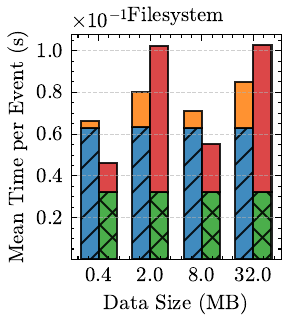}
        \caption{8 nodes}
        \label{fig:512_nodes}
    \end{subfigure}
    \begin{subfigure}[b]{0.48\linewidth}
        \centering
        \includegraphics[width=\linewidth]{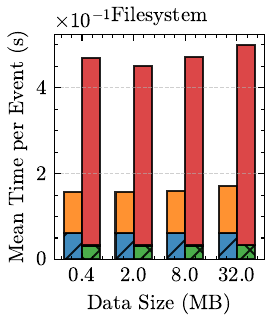}
        \caption{512 nodes}
        \label{fig:bottom_right}
    \end{subfigure}

    \caption{Comparison of time spent in computation (AI iter, Sim iter) versus data transport (read, write). The analysis covers the node-local backend (top row) and the file system backend (bottom row) at 8-node (left column) and 512-node (right column) scales. 
    }
    \label{fig:fig4_nekrs_ml_run_time_comparison_grid}

    \Description{
    This figure contains four stacked bar charts arranged in a two-by-two grid, comparing two mechanisms at two different scales. The top row, labeled 'Nodelocal', shows performance at 8 nodes (a) and 512 nodes (b). The bottom row, labeled 'Filesystem', shows the same comparison for 8 nodes (c) and 512 nodes (d). For all charts, the horizontal axis represents four 'Data Size' values in megabytes, and the vertical axis represents the 'Mean Time per Event' in seconds. The vertical axis scale for the top two Nodelocal charts is multiplied by ten to the power of negative two, while the scale for the bottom two Filesystem charts is multiplied by ten to the power of negative one. Each bar in the charts is a stack of four segments representing the time contributions from four event types: 'AI iter', 'read', 'Sim iter', and 'write'. The height of each segment corresponds to the time for that specific event, and the total height of the stacked bar represents the total mean time per event for a given data size and configuration.    
    }
\end{figure}

To show the importance of optimizing data transport, Fig. \ref{fig:fig4_nekrs_ml_run_time_comparison_grid} compares the mean computation time (iter) against the data transport time per message. We present the node-local and file system backends as they represent the extremes of the scaling performance observed in our tests. For both backends, the data transport time generally increases with message size, but their overhead relative to the computation time is significantly different.

The node-local backend demonstrates very low overhead. Even at the largest message size of 32 MB, the time for a single data transfer is roughly equal to one computation iteration. Since data is written only once every 100 iterations in this workflow, this represents a negligible performance cost. Furthermore, the node-local backend scales extremely well, showing almost no change in data transport time when scaling from 8 to 512 nodes.

In contrast, the file system overhead grows dramatically with scale. While it performs reasonably well at 8 nodes, where a 32 MB transfer takes approximately one iteration time, its performance degrades at 512 nodes. At this larger scale, the data transport creates a significant overhead, with the transfer time becoming approximately an order of magnitude larger than one iteration. This highlights that poor choice of backend can cause a workflow performance to be completely dominated by the data transport between components.

\subsection{Pattern 2: Many-to-One}

This workflow pattern represents a common use case for online training of surrogate models, where one AI model is trained from a large ensemble of concurrent simulations \cite{breweraicoupled2024}. Unlike the previous co-located pattern, the simulation components and the single AI training component run on different nodes. We expect the required non-local data transport to be a significant bottleneck, and the following experiments are designed to test how each of the backends perform and scale under these conditions. 
Moreover, while this section does not aim to validate a mini-app against a realistic workflow, it highlights how SimAI-Bench can be used to benchmark and prototype new patterns and inform implementation choices.

The experiment is designed as follows: each simulation component writes a data array every 10 iterations. The AI component, in turn, reads the data from all simulations every 10 iterations. It is important to note that the AI component blocks until all data for that specific update iteration has arrived, ensuring a consistent workload for each test regardless of the backend read speed. Each component is allocated its own node and utilizes all 12 GPU tiles on Aurora. The iteration times required for AI and simulation components of the mini-app are obtained from the same applications in Pattern 1. 

\begin{figure}[htbp]
    \centering
    \begin{subfigure}[b]{0.48\linewidth}
        \centering
        \includegraphics[width=\linewidth]{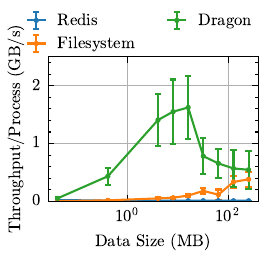}
        \caption{Read}
    \end{subfigure}
    \hfill
    \begin{subfigure}[b]{0.48\linewidth}
        \centering
        \includegraphics[width=\linewidth]{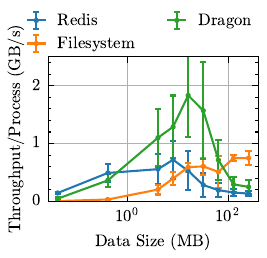}
        \caption{Write}
    \end{subfigure}

    \caption{Throughput for the 2-node implementation of  Pattern 2, showing (a) non-local read and (b) local write performance as a function of array size. 
    }
    \label{fig:fig5_msr_rw_throughput}
    \Description{
    This figure consists of two line plots, labeled (a) Read and (b) Write. Both plots have a horizontal x-axis representing 'Data Size' in megabytes on a logarithmic scale and a vertical y-axis representing 'Throughput/Process' in gigabytes per second on a linear scale. Each plot displays three lines corresponding to different systems: Redis, Filesystem, and Dragon. Each line connects a series of data points that show the measured throughput at various data sizes. Vertical error bars are shown for each data point.
    }
\end{figure}

Before scaling the many-to-one pattern, our first experiment uses a 2-node workflow, with the simulation and AI training components on different nodes. The simulation stages data to its local backend, and the AI component reads this data non-locally. The opposite case (non-local write, local read) showed similar results and is not presented here. 
Of course, when using the file system backend, the I/O is performed to the disk and not locally to a node. Additionally, a node-local solution using tmpfs is not possible in this case and thus was excluded from the backends tested.
Fig. \ref{fig:fig5_msr_rw_throughput} shows the read and write throughput from this experiment. The data presented here is averaged from all the processes and events.

The results highlight the performance challenge of non-local data access. Redis, for instance, performs poorly on the non-local read task, even though its local write performance is reasonable. In contrast, DragonHPC shows high throughput for both local writes and non-local reads, though its performance peaks at a message size of approximately 10 MB before declining. The file system throughput increases continuously with data size, similar to its behavior in Pattern 1, eventually becoming comparable to DragonHPC at the largest message sizes. Notably, the performance profiles of the local writes for this pattern is similar to Fig. \ref{fig:fig3_nekrs_ml_rw_throughput}.

\begin{figure}[htbp]
    \centering
    \begin{subfigure}[b]{0.48\linewidth}
        \centering
        \includegraphics[width=\linewidth]{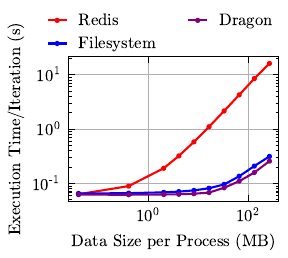}
        \caption{8 nodes}
    \end{subfigure}
    \begin{subfigure}[b]{0.48\linewidth}
        \centering
        \includegraphics[width=\linewidth]{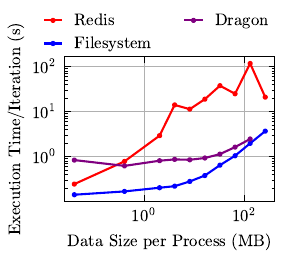}
        \caption{128 nodes}
    \end{subfigure}

    \caption{Training runtime per iteration for Pattern 2 as a function of data size at (a) 8 nodes and (b) 128 nodes. 
    }
    \Description{
    This figure shows two line plots, labeled (a) 8 nodes and (b) 128 nodes. Both plots use a log-log scale, with the vertical y-axis representing 'Execution Time/Iteration' in seconds and the horizontal x-axis representing 'Data Size per Process' in megabytes. Each plot contains three lines with circular markers, each corresponding to a different data transfer system: Redis (red), Filesystem (blue), and Dragon (purple). The lines trace the execution time as a function of data size for each system under the respective node count.
    }
    \label{fig:fig6_msr_training_runtime}
\end{figure}

In our second experiment, we scaled this local-write and non-local-read pattern by increasing the number of simulation components proportionally with the node count (one per node), while keeping a single AI component. Fig. \ref{fig:fig6_msr_training_runtime} shows the resulting training runtime per iteration at various data sizes and node counts. Here, execution time per iteration is obtained by computing the total execution time of the training component divided by the number of iterations. Hence, this includes both compute and data transport times.

At 8 nodes, the execution time for all backends increases with data size. The execution time for Redis increases most significantly, which is expected given its low non-local read throughput. At this scale, the DragonHPC and file system backends perform equally well.

As we scale up to 128 nodes, the performance characteristics change. While Redis remains the slowest, the relative performance of DragonHPC and the file system is highly dependent on the message size. For message sizes less than 10 MB, DragonHPC runtime is significantly longer than the file system. Considering that DragonHPC's point-to-point throughput peaks in this range with significantly larger throughput than the file system, this difference points to increased latency when handling many-to-one communication. For larger message sizes, both DragonHPC and the file system show similar performance.

From the above analysis it is clear that using file system is currently the most optimal backend for this pattern.

\section{Conclusions and Future Work}
\label{sec:conclusion}
The increasing complexity and data-intensive nature of coupled AI-simulation workflows necessitate new tools for performance analysis and optimization. In this paper, we introduced SimAI-Bench, a flexible Python API for constructing workflow mini-apps designed to benchmark and optimize the critical data transport patterns at the core of such workflows. We demonstrated its effectiveness by modeling two common workflow patterns and evaluating the performance of various data staging backends.

Our evaluation on the Aurora supercomputer yielded several key insights. For co-located, one-to-one workflows, we showed that node-local and DragonHPC distributed dictionary data stores provide excellent and scalable performance, while shared parallel file systems can suffer from severe metadata contention at high node counts. For distributed, many-to-one ensemble workflows, our findings revealed that file system performs better than the in-memory datastores. We found that high point-to-point throughput does not always guarantee the best performance in a many-to-one communication pattern, suggesting that latency can become a critical factor.

These results underscore that a one-size-fits-all approach to data transport is insufficient for modern scientific workflows. The ability to prototype and benchmark different strategies, as enabled by SimAI-Bench, is crucial for application developers and facility operators to achieve optimal performance. 
Furthermore, we acknowledge that some key insights obtained may be specific to the hardware on the Aurora supercomputer. In particular, while we expect the node-local solution for co-located, one-to-one workflows to be the most performant on most systems due to its high memory bandwidth and reduced overhead from third party staging libraries, the best data transport strategy for distributed, many-to-one workflows may vary depending on the available interconnect and file system. This fact highlights the portability challenges faced by developers of AI-coupled workflows, which SimAI-Bench aims to solve by providing portable Python mini-apps for easy benchmarking and prototyping. 

Future work will focus on increasing the fidelity of our component emulation by incorporating more detailed models for communication and arithmetic intensity of computational kernels. We also plan to expand the library of emulated AI components to support more advanced model architectures, such as the Graph Neural Networks used in our target science case for Pattern 1. Additionally, we plan add support for point-to-point streaming, for instance using ADIOS2, and for staging through DAOS on Aurora. 
Finally, while SimAI-Bench is already capable of modeling full end-to-end workflows, we intend to expand our benchmarking analysis to tackle more complex AI-coupled applications on multiple HPC systems, which will require interfacing SimAI-Bench with workflow orchestration tools such as RADICAL-Pilot.

\begin{acks}
This research used resources of the Argonne Leadership Computing Facility, a U.S. Department of Energy (DOE) Office of Science user facility at Argonne National Laboratory and is based on research supported by the U.S. DOE Office of Science-Advanced Scientific Computing Research Program, under Contract No. DE-AC02-06CH11357.
\end{acks}

\bibliographystyle{ACM-Reference-Format}
\bibliography{references}


\begin{thebibliography}{30}


\ifx \showCODEN    \undefined \def \showCODEN     #1{\unskip}     \fi
\ifx \showISBNx    \undefined \def \showISBNx     #1{\unskip}     \fi
\ifx \showISBNxiii \undefined \def \showISBNxiii  #1{\unskip}     \fi
\ifx \showISSN     \undefined \def \showISSN      #1{\unskip}     \fi
\ifx \showLCCN     \undefined \def \showLCCN      #1{\unskip}     \fi
\ifx \shownote     \undefined \def \shownote      #1{#1}          \fi
\ifx \showarticletitle \undefined \def \showarticletitle #1{#1}   \fi
\ifx \showURL      \undefined \def \showURL       {\relax}        \fi
\providecommand\bibfield[2]{#2}
\providecommand\bibinfo[2]{#2}
\providecommand\natexlab[1]{#1}
\providecommand\showeprint[2][]{arXiv:#2}

\bibitem[Applencourt et~al\mbox{.}(2024)]%
        {pvc_paper}
\bibfield{author}{\bibinfo{person}{Thomas Applencourt}, \bibinfo{person}{Aditya Sadawarte}, \bibinfo{person}{Servesh Muralidharan}, \bibinfo{person}{Colleen Bertoni}, \bibinfo{person}{JaeHyuk Kwack}, \bibinfo{person}{Ye Luo}, \bibinfo{person}{Esteban Rangel}, \bibinfo{person}{John Tramm}, \bibinfo{person}{Yasaman Ghadar}, \bibinfo{person}{Arjen Tamerus}, \bibinfo{person}{Chris Edsall}, {and} \bibinfo{person}{Tom Deakin}.} \bibinfo{year}{2024}\natexlab{}.
\newblock \showarticletitle{Ponte Vecchio Across the Atlantic: Single-Node Benchmarking of Two Intel GPU Systems}. In \bibinfo{booktitle}{\emph{SC24-W: Workshops of the International Conference for High Performance Computing, Networking, Storage and Analysis}}. \bibinfo{pages}{1430--1442}.
\newblock
\href{https://doi.org/10.1109/SCW63240.2024.00184}{doi:\nolinkurl{10.1109/SCW63240.2024.00184}}


\bibitem[Balin et~al\mbox{.}(2023)]%
        {balin2023situ}
\bibfield{author}{\bibinfo{person}{Riccardo Balin}, \bibinfo{person}{Filippo Simini}, \bibinfo{person}{Cooper Simpson}, \bibinfo{person}{Andrew Shao}, \bibinfo{person}{Alessandro Rigazzi}, \bibinfo{person}{Matthew Ellis}, \bibinfo{person}{Stephen Becker}, \bibinfo{person}{Alireza Doostan}, \bibinfo{person}{John~A. Evans}, {and} \bibinfo{person}{Kenneth~E. Jansen}.} \bibinfo{year}{2023}\natexlab{}.
\newblock \bibinfo{title}{In Situ Framework for Coupling Simulation and Machine Learning with Application to CFD}.
\newblock
\showeprint[arxiv]{2306.12900}~[cs.LG]
\urldef\tempurl%
\url{https://arxiv.org/abs/2306.12900}
\showURL{%
\tempurl}


\bibitem[Bard et~al\mbox{.}(2023)]%
        {bard2023workflow}
\bibfield{author}{\bibinfo{person}{Deborah Bard}, \bibinfo{person}{Taylor Groves}, \bibinfo{person}{Brandon Cook}, \bibinfo{person}{Laurie Stephey}, \bibinfo{person}{Wahid Bhimji}, \bibinfo{person}{Steve Farrell}, \bibinfo{person}{Brian Austin}, \bibinfo{person}{Kevin Gott}, \bibinfo{person}{Shane Canon}, \bibinfo{person}{Kristy Kallback-Rose}, {et~al\mbox{.}}} \bibinfo{year}{2023}\natexlab{}.
\newblock \bibinfo{title}{Workflow Archetypes White Paper}.
\newblock


\bibitem[Barrett et~al\mbox{.}(2015)]%
        {barrett_assessing_2015}
\bibfield{author}{\bibinfo{person}{R.~F. Barrett}, \bibinfo{person}{P.~S. Crozier}, \bibinfo{person}{D.~W. Doerfler}, \bibinfo{person}{M.~A. Heroux}, \bibinfo{person}{P.~T. Lin}, \bibinfo{person}{H.~K. Thornquist}, \bibinfo{person}{T.~G. Trucano}, {and} \bibinfo{person}{C.~T. Vaughan}.} \bibinfo{year}{2015}\natexlab{}.
\newblock \showarticletitle{Assessing the role of mini-applications in predicting key performance characteristics of scientific and engineering applications}.
\newblock \bibinfo{journal}{\emph{J. Parallel and Distrib. Comput.}}  \bibinfo{volume}{75} (\bibinfo{date}{Jan.} \bibinfo{year}{2015}), \bibinfo{pages}{107--122}.
\newblock
\showISSN{0743-7315}
\href{https://doi.org/10.1016/j.jpdc.2014.09.006}{doi:\nolinkurl{10.1016/j.jpdc.2014.09.006}}


\bibitem[Barwey et~al\mbox{.}(2024)]%
        {balin_gnn}
\bibfield{author}{\bibinfo{person}{Shivam Barwey}, \bibinfo{person}{Riccardo Balin}, \bibinfo{person}{Bethany Lusch}, \bibinfo{person}{Saumil Patel}, \bibinfo{person}{Ramesh Balakrishnan}, \bibinfo{person}{Pinaki Pal}, \bibinfo{person}{Romit Maulik}, {and} \bibinfo{person}{Venkatram Vishwanath}.} \bibinfo{year}{2024}\natexlab{}.
\newblock \showarticletitle{Scalable and Consistent Graph Neural Networks for Distributed Mesh-based Data-driven Modeling}. In \bibinfo{booktitle}{\emph{SC24-W: Workshops of the International Conference for High Performance Computing, Networking, Storage and Analysis}}. \bibinfo{pages}{1058--1070}.
\newblock
\href{https://doi.org/10.1109/SCW63240.2024.00146}{doi:\nolinkurl{10.1109/SCW63240.2024.00146}}


\bibitem[Boggia et~al\mbox{.}(2025)]%
        {boggia2025review}
\bibfield{author}{\bibinfo{person}{Laura Boggia}, \bibinfo{person}{Carlos Cocha}, \bibinfo{person}{Fotis Giasemis}, \bibinfo{person}{Joachim Hansen}, \bibinfo{person}{Patin Inkaew}, \bibinfo{person}{Kaare~Endrup Iversen}, \bibinfo{person}{Pratik Jawahar}, \bibinfo{person}{Henrique~Pineiro Monteagudo}, \bibinfo{person}{Micol Olocco}, \bibinfo{person}{Sten Astrand}, \bibinfo{person}{Martino Borsato}, \bibinfo{person}{Leon Bozianu}, \bibinfo{person}{Steven Schramm}, {and} \bibinfo{person}{the SMARTHEP~Network}.} \bibinfo{year}{2025}\natexlab{}.
\newblock \bibinfo{title}{Review of Machine Learning for Real-Time Analysis at the Large Hadron Collider experiments ALICE, ATLAS, CMS and LHCb}.
\newblock
\showeprint[arxiv]{2506.14578}~[hep-ex]
\urldef\tempurl%
\url{https://arxiv.org/abs/2506.14578}
\showURL{%
\tempurl}


\bibitem[Brewer et~al\mbox{.}(2024)]%
        {breweraicoupled2024}
\bibfield{author}{\bibinfo{person}{Wes Brewer}, \bibinfo{person}{Ana Gainaru}, \bibinfo{person}{Frédéric Suter}, \bibinfo{person}{Feiyi Wang}, \bibinfo{person}{Murali Emani}, {and} \bibinfo{person}{Shantenu Jha}.} \bibinfo{year}{2024}\natexlab{}.
\newblock \bibinfo{title}{{AI}-coupled {HPC} {Workflow} {Applications}, {Middleware} and {Performance}}.
\newblock
\href{https://doi.org/10.48550/arXiv.2406.14315}{doi:\nolinkurl{10.48550/arXiv.2406.14315}}
\newblock
\shownote{arXiv:2406.14315 [cs]}.


\bibitem[Coleman et~al\mbox{.}(2021)]%
        {coleman2021wfchef}
\bibfield{author}{\bibinfo{person}{Tainã Coleman}, \bibinfo{person}{Henri Casanova}, {and} \bibinfo{person}{Rafael~Ferreira da Silva}.} \bibinfo{year}{2021}\natexlab{}.
\newblock \showarticletitle{WfChef: Automated Generation of Accurate Scientific Workflow Generators}. In \bibinfo{booktitle}{\emph{2021 IEEE 17th International Conference on eScience (eScience)}}. \bibinfo{pages}{159--168}.
\newblock
\href{https://doi.org/10.1109/eScience51609.2021.00026}{doi:\nolinkurl{10.1109/eScience51609.2021.00026}}


\bibitem[Coleman et~al\mbox{.}(2022)]%
        {coleman2022wfbench}
\bibfield{author}{\bibinfo{person}{Taina Coleman}, \bibinfo{person}{Henri Casanova}, \bibinfo{person}{Ketan Maheshwari}, \bibinfo{person}{Loic Pottier}, \bibinfo{person}{Sean~R. Wilkinson}, \bibinfo{person}{Justin Wozniak}, \bibinfo{person}{Frederic Suter}, \bibinfo{person}{Mallikarjun Shankar}, {and} \bibinfo{person}{Rafael~Ferreira Da~Silva}.} \bibinfo{year}{2022}\natexlab{}.
\newblock \showarticletitle{{ WfBench: Automated Generation of Scientific Workflow Benchmarks }}. In \bibinfo{booktitle}{\emph{2022 IEEE/ACM International Workshop on Performance Modeling, Benchmarking and Simulation of High Performance Computer Systems (PMBS)}}. \bibinfo{publisher}{IEEE Computer Society}, \bibinfo{address}{Los Alamitos, CA, USA}, \bibinfo{pages}{100--111}.
\newblock
\href{https://doi.org/10.1109/PMBS56514.2022.00014}{doi:\nolinkurl{10.1109/PMBS56514.2022.00014}}


\bibitem[{Computational Data Analysis Workflow Systems}({[n.\,d.]})]%
        {workflow_list}
\bibfield{author}{\bibinfo{person}{{Computational Data Analysis Workflow Systems}}.} \bibinfo{year}{[n.\,d.]}\natexlab{}.
\newblock \bibinfo{howpublished}{\url{https://s.apache.org/existing-workflow-systems}}.
\newblock
\newblock
\shownote{Accessed: 2025-08-03}.


\bibitem[Crozier et~al\mbox{.}(2009)]%
        {crozier_improving_2009}
\bibfield{author}{\bibinfo{person}{Paul Crozier}, \bibinfo{person}{Heidi Thornquist}, \bibinfo{person}{Robert Numrich}, \bibinfo{person}{Alan Williams}, \bibinfo{person}{Harold Edwards}, \bibinfo{person}{Eric Keiter}, \bibinfo{person}{Mahesh Rajan}, \bibinfo{person}{James Willenbring}, \bibinfo{person}{Douglas Doerfler}, {and} \bibinfo{person}{Michael Heroux}.} \bibinfo{year}{2009}\natexlab{}.
\newblock \bibinfo{booktitle}{\emph{Improving performance via mini-applications.}}
\newblock \bibinfo{type}{{T}echnical {R}eport} SAND2009-5574, 993908. \bibinfo{pages}{SAND2009--5574, 993908} pages.
\newblock
\href{https://doi.org/10.2172/993908}{doi:\nolinkurl{10.2172/993908}}


\bibitem[Docan et~al\mbox{.}(2010)]%
        {docan2010dataspaces}
\bibfield{author}{\bibinfo{person}{Ciprian Docan}, \bibinfo{person}{Manish Parashar}, {and} \bibinfo{person}{Scott Klasky}.} \bibinfo{year}{2010}\natexlab{}.
\newblock \showarticletitle{DataSpaces: an interaction and coordination framework for coupled simulation workflows}. In \bibinfo{booktitle}{\emph{Proceedings of the 19th ACM International Symposium on High Performance Distributed Computing}} (Chicago, Illinois) \emph{(\bibinfo{series}{HPDC '10})}. \bibinfo{publisher}{Association for Computing Machinery}, \bibinfo{address}{New York, NY, USA}, \bibinfo{pages}{25–36}.
\newblock
\showISBNx{9781605589428}
\href{https://doi.org/10.1145/1851476.1851481}{doi:\nolinkurl{10.1145/1851476.1851481}}


\bibitem[{dpnp}({[n.\,d.]})]%
        {dpnp}
\bibfield{author}{\bibinfo{person}{{dpnp}}.} \bibinfo{year}{[n.\,d.]}\natexlab{}.
\newblock \bibinfo{howpublished}{\url{https://github.com/IntelPython/dpnp}}.
\newblock
\newblock
\shownote{Accessed: 2025-08-07}.


\bibitem[{DragonHPC}({[n.\,d.]})]%
        {DragonHPC}
\bibfield{author}{\bibinfo{person}{{DragonHPC}}.} \bibinfo{year}{[n.\,d.]}\natexlab{}.
\newblock \bibinfo{howpublished}{\url{https://dragonhpc.org/portal/index.html}}.
\newblock
\newblock
\shownote{Accessed: 2025-08-03}.


\bibitem[Fischer et~al\mbox{.}(2022)]%
        {fischer2022nekrs}
\bibfield{author}{\bibinfo{person}{Paul Fischer}, \bibinfo{person}{Stefan Kerkemeier}, \bibinfo{person}{Misun Min}, \bibinfo{person}{Yu-Hsiang Lan}, \bibinfo{person}{Malachi Phillips}, \bibinfo{person}{Thilina Rathnayake}, \bibinfo{person}{Elia Merzari}, \bibinfo{person}{Ananias Tomboulides}, \bibinfo{person}{Ali Karakus}, \bibinfo{person}{Noel Chalmers}, {and} \bibinfo{person}{Tim Warburton}.} \bibinfo{year}{2022}\natexlab{}.
\newblock \showarticletitle{NekRS, a GPU-accelerated spectral element Navier–Stokes solver}.
\newblock \bibinfo{journal}{\emph{Parallel Comput.}}  \bibinfo{volume}{114} (\bibinfo{year}{2022}), \bibinfo{pages}{102982}.
\newblock
\showISSN{0167-8191}
\href{https://doi.org/10.1016/j.parco.2022.102982}{doi:\nolinkurl{10.1016/j.parco.2022.102982}}


\bibitem[Godoy et~al\mbox{.}(2020)]%
        {godoy2020adios}
\bibfield{author}{\bibinfo{person}{William~F. Godoy}, \bibinfo{person}{Norbert Podhorszki}, \bibinfo{person}{Ruonan Wang}, \bibinfo{person}{Chuck Atkins}, \bibinfo{person}{Greg Eisenhauer}, \bibinfo{person}{Junmin Gu}, \bibinfo{person}{Philip Davis}, \bibinfo{person}{Jong Choi}, \bibinfo{person}{Kai Germaschewski}, \bibinfo{person}{Kevin Huck}, \bibinfo{person}{Axel Huebl}, \bibinfo{person}{Mark Kim}, \bibinfo{person}{James Kress}, \bibinfo{person}{Tahsin Kurc}, \bibinfo{person}{Qing Liu}, \bibinfo{person}{Jeremy Logan}, \bibinfo{person}{Kshitij Mehta}, \bibinfo{person}{George Ostrouchov}, \bibinfo{person}{Manish Parashar}, \bibinfo{person}{Franz Poeschel}, \bibinfo{person}{David Pugmire}, \bibinfo{person}{Eric Suchyta}, \bibinfo{person}{Keichi Takahashi}, \bibinfo{person}{Nick Thompson}, \bibinfo{person}{Seiji Tsutsumi}, \bibinfo{person}{Lipeng Wan}, \bibinfo{person}{Matthew Wolf}, \bibinfo{person}{Kesheng Wu}, {and} \bibinfo{person}{Scott Klasky}.} \bibinfo{year}{2020}\natexlab{}.
\newblock \showarticletitle{ADIOS 2: The Adaptable Input Output System. A framework for high-performance data management}.
\newblock \bibinfo{journal}{\emph{SoftwareX}}  \bibinfo{volume}{12} (\bibinfo{year}{2020}), \bibinfo{pages}{100561}.
\newblock
\showISSN{2352-7110}
\href{https://doi.org/10.1016/j.softx.2020.100561}{doi:\nolinkurl{10.1016/j.softx.2020.100561}}


\bibitem[Jha et~al\mbox{.}({[n.\,d.]})]%
        {jha2022ai_v2}
\bibfield{author}{\bibinfo{person}{Shantenu Jha}, \bibinfo{person}{Vincent Pascuzzi}, {and} \bibinfo{person}{Matteo Turilli}.} \bibinfo{year}{[n.\,d.]}\natexlab{}.
\newblock \bibinfo{booktitle}{\emph{AI-coupled HPC Workflows}}.
\newblock Chapter Chapter 28, \bibinfo{pages}{515--534}.
\newblock
\showeprint{https://www.worldscientific.com/doi/pdf/10.1142/9789811265679\_0028}
\href{https://doi.org/10.1142/9789811265679\_0028}{doi:\nolinkurl{10.1142/9789811265679\_0028}}


\bibitem[Kilic et~al\mbox{.}(2024)]%
        {kilic_workflow_2024}
\bibfield{author}{\bibinfo{person}{Ozgur~O. Kilic}, \bibinfo{person}{Tianle Wang}, \bibinfo{person}{Matteo Turilli}, \bibinfo{person}{Mikhail Titov}, \bibinfo{person}{Andre Merzky}, \bibinfo{person}{Line Pouchard}, {and} \bibinfo{person}{Shantenu Jha}.} \bibinfo{year}{2024}\natexlab{}.
\newblock \showarticletitle{Workflow {Mini}-{Apps}: {Portable}, {Scalable}, {Tunable} \& {Faithful} {Representations} of {Scientific} {Workflows}}. In \bibinfo{booktitle}{\emph{2024 {IEEE} 24th {International} {Symposium} on {Cluster}, {Cloud} and {Internet} {Computing} ({CCGrid})}}. \bibinfo{pages}{465--477}.
\newblock
\href{https://doi.org/10.1109/CCGrid59990.2024.00059}{doi:\nolinkurl{10.1109/CCGrid59990.2024.00059}}
\newblock
\shownote{ISSN: 2993-2114}.


\bibitem[Kwack et~al\mbox{.}(2021)]%
        {kwack_evaluation_2021}
\bibfield{author}{\bibinfo{person}{JaeHyuk Kwack}, \bibinfo{person}{John Tramm}, \bibinfo{person}{Colleen Bertoni}, \bibinfo{person}{Yasaman Ghadar}, \bibinfo{person}{Brian Homerding}, \bibinfo{person}{Esteban Rangel}, \bibinfo{person}{Christopher Knight}, {and} \bibinfo{person}{Scott Parker}.} \bibinfo{year}{2021}\natexlab{}.
\newblock \showarticletitle{Evaluation of {Performance} {Portability} of {Applications} and {Mini}-{Apps} across {AMD}, {Intel} and {NVIDIA} {GPUs}}. In \bibinfo{booktitle}{\emph{2021 {International} {Workshop} on {Performance}, {Portability} and {Productivity} in {HPC} ({P3HPC})}}. \bibinfo{pages}{45--56}.
\newblock
\href{https://doi.org/10.1109/P3HPC54578.2021.00008}{doi:\nolinkurl{10.1109/P3HPC54578.2021.00008}}


\bibitem[Lee et~al\mbox{.}(2019)]%
        {lee2019deepdrivemd}
\bibfield{author}{\bibinfo{person}{Hyungro Lee}, \bibinfo{person}{Matteo Turilli}, \bibinfo{person}{Shantenu Jha}, \bibinfo{person}{Debsindhu Bhowmik}, \bibinfo{person}{Heng Ma}, {and} \bibinfo{person}{Arvind Ramanathan}.} \bibinfo{year}{2019}\natexlab{}.
\newblock \showarticletitle{{ DeepDriveMD: Deep-Learning Driven Adaptive Molecular Simulations for Protein Folding }}. In \bibinfo{booktitle}{\emph{2019 IEEE/ACM Third Workshop on Deep Learning on Supercomputers (DLS)}}. \bibinfo{publisher}{IEEE Computer Society}, \bibinfo{address}{Los Alamitos, CA, USA}, \bibinfo{pages}{12--19}.
\newblock
\href{https://doi.org/10.1109/DLS49591.2019.00007}{doi:\nolinkurl{10.1109/DLS49591.2019.00007}}


\bibitem[Luszczek et~al\mbox{.}(2006)]%
        {luszczek2006hpc}
\bibfield{author}{\bibinfo{person}{Piotr~R Luszczek}, \bibinfo{person}{David~H Bailey}, \bibinfo{person}{Jack~J Dongarra}, \bibinfo{person}{Jeremy Kepner}, \bibinfo{person}{Robert~F Lucas}, \bibinfo{person}{Rolf Rabenseifner}, {and} \bibinfo{person}{Daisuke Takahashi}.} \bibinfo{year}{2006}\natexlab{}.
\newblock \showarticletitle{The HPC Challenge (HPCC) benchmark suite}. In \bibinfo{booktitle}{\emph{Proceedings of the 2006 ACM/IEEE Conference on Supercomputing}} (Tampa, Florida) \emph{(\bibinfo{series}{SC '06})}. \bibinfo{publisher}{Association for Computing Machinery}, \bibinfo{address}{New York, NY, USA}, \bibinfo{pages}{213–es}.
\newblock
\showISBNx{0769527000}
\href{https://doi.org/10.1145/1188455.1188677}{doi:\nolinkurl{10.1145/1188455.1188677}}


\bibitem[Martineau and McIntosh-Smith(2017)]%
        {martineau_arch_2017}
\bibfield{author}{\bibinfo{person}{Matthew Martineau} {and} \bibinfo{person}{Simon McIntosh-Smith}.} \bibinfo{year}{2017}\natexlab{}.
\newblock \showarticletitle{The {Arch} {Project}: {Physics} {Mini}-{Apps} for {Algorithmic} {Exploration} and {Evaluating} {Programming} {Environments} on {HPC} {Architectures}}. In \bibinfo{booktitle}{\emph{2017 {IEEE} {International} {Conference} on {Cluster} {Computing} ({CLUSTER})}}. \bibinfo{pages}{850--857}.
\newblock
\href{https://doi.org/10.1109/CLUSTER.2017.126}{doi:\nolinkurl{10.1109/CLUSTER.2017.126}}
\newblock
\shownote{ISSN: 2168-9253}.


\bibitem[Marts et~al\mbox{.}(2021)]%
        {marts_minimod_2021}
\bibfield{author}{\bibinfo{person}{W.~Pepper Marts}, \bibinfo{person}{Matthew G.~F. Dosanjh}, \bibinfo{person}{Scott Levy}, \bibinfo{person}{Whit Schonbein}, \bibinfo{person}{Ryan~E. Grant}, {and} \bibinfo{person}{Patrick~G. Bridges}.} \bibinfo{year}{2021}\natexlab{}.
\newblock \showarticletitle{{MiniMod}: {A} {Modular} {Miniapplication} {Benchmarking} {Framework} for {HPC}}. In \bibinfo{booktitle}{\emph{2021 {IEEE} {International} {Conference} on {Cluster} {Computing} ({CLUSTER})}}. \bibinfo{pages}{12--22}.
\newblock
\href{https://doi.org/10.1109/Cluster48925.2021.00028}{doi:\nolinkurl{10.1109/Cluster48925.2021.00028}}
\newblock
\shownote{ISSN: 2168-9253}.


\bibitem[Mas~Magre et~al\mbox{.}(2025)]%
        {mas_magre_nomad_2025}
\bibfield{author}{\bibinfo{person}{Isidre Mas~Magre}, \bibinfo{person}{Rogeli Grima~Torres}, \bibinfo{person}{José~María Cela~Espín}, {and} \bibinfo{person}{José~Julio Gutierrez~Moreno}.} \bibinfo{year}{2025}\natexlab{}.
\newblock \showarticletitle{The {NOMAD} mini-apps: {A} suite of kernels from ab initio electronic structure codes enabling co-design in high-performance computing}.
\newblock \bibinfo{journal}{\emph{Open Research Europe}}  \bibinfo{volume}{4} (\bibinfo{date}{April} \bibinfo{year}{2025}), \bibinfo{pages}{35}.
\newblock
\showISSN{2732-5121}
\href{https://doi.org/10.12688/openreseurope.16920.3}{doi:\nolinkurl{10.12688/openreseurope.16920.3}}


\bibitem[Messer et~al\mbox{.}(2018)]%
        {messer_miniapps_2018}
\bibfield{author}{\bibinfo{person}{OE~Bronson Messer}, \bibinfo{person}{Ed D’Azevedo}, \bibinfo{person}{Judy Hill}, \bibinfo{person}{Wayne Joubert}, \bibinfo{person}{Mark Berrill}, {and} \bibinfo{person}{Christopher Zimmer}.} \bibinfo{year}{2018}\natexlab{}.
\newblock \showarticletitle{{MiniApps} derived from production {HPC} applications using multiple programing models}.
\newblock \bibinfo{journal}{\emph{The International Journal of High Performance Computing Applications}} \bibinfo{volume}{32}, \bibinfo{number}{4} (\bibinfo{date}{July} \bibinfo{year}{2018}), \bibinfo{pages}{582--593}.
\newblock
\showISSN{1094-3420}
\href{https://doi.org/10.1177/1094342016668241}{doi:\nolinkurl{10.1177/1094342016668241}}
\newblock
\shownote{Publisher: SAGE Publications Ltd STM}.


\bibitem[Partee et~al\mbox{.}(2022)]%
        {partee2022}
\bibfield{author}{\bibinfo{person}{Sam Partee}, \bibinfo{person}{Matthew Ellis}, \bibinfo{person}{Alessandro Rigazzi}, \bibinfo{person}{Andrew~E. Shao}, \bibinfo{person}{Scott Bachman}, \bibinfo{person}{Gustavo Marques}, {and} \bibinfo{person}{Benjamin Robbins}.} \bibinfo{year}{2022}\natexlab{}.
\newblock \showarticletitle{Using Machine Learning at scale in numerical simulations with SmartSim: An application to ocean climate modeling}.
\newblock \bibinfo{journal}{\emph{Journal of Computational Science}}  \bibinfo{volume}{62} (\bibinfo{year}{2022}), \bibinfo{pages}{101707}.
\newblock
\showISSN{1877-7503}
\href{https://doi.org/10.1016/j.jocs.2022.101707}{doi:\nolinkurl{10.1016/j.jocs.2022.101707}}


\bibitem[Saurabh et~al\mbox{.}(2024)]%
        {saurabh_quantum_2024}
\bibfield{author}{\bibinfo{person}{Nishant Saurabh}, \bibinfo{person}{Pradeep Mantha}, \bibinfo{person}{Florian~J. Kiwit}, \bibinfo{person}{Shantenu Jha}, {and} \bibinfo{person}{Andre Luckow}.} \bibinfo{year}{2024}\natexlab{}.
\newblock \showarticletitle{Quantum {Mini}-{Apps}: {A} {Framework} for {Developing} and {Benchmarking} {Quantum}-{HPC} {Applications}}. In \bibinfo{booktitle}{\emph{Proceedings of the 2024 {Workshop} on {High} {Performance} and {Quantum} {Computing} {Integration}}} \emph{(\bibinfo{series}{{HPQCI} '24})}. \bibinfo{publisher}{Association for Computing Machinery}, \bibinfo{address}{New York, NY, USA}, \bibinfo{pages}{11--18}.
\newblock
\showISBNx{979-8-4007-0643-1}
\href{https://doi.org/10.1145/3659996.3660036}{doi:\nolinkurl{10.1145/3659996.3660036}}


\bibitem[{SimAI-Bench}({[n.\,d.]})]%
        {simaibench}
\bibfield{author}{\bibinfo{person}{{SimAI-Bench}}.} \bibinfo{year}{[n.\,d.]}\natexlab{}.
\newblock \bibinfo{howpublished}{\url{https://github.com/argonne-lcf/SimAI-Bench}}.
\newblock
\newblock
\shownote{Accessed: 2025-08-08}.


\bibitem[Sukumar et~al\mbox{.}(2016)]%
        {sukumar_mini-apps_2016}
\bibfield{author}{\bibinfo{person}{Sreenivas~R. Sukumar}, \bibinfo{person}{Michael~A. Matheson}, \bibinfo{person}{Ramakrishnan Kannan}, {and} \bibinfo{person}{Seung-Hwan Lim}.} \bibinfo{year}{2016}\natexlab{}.
\newblock \showarticletitle{Mini-apps for high performance data analysis}. In \bibinfo{booktitle}{\emph{2016 {IEEE} {International} {Conference} on {Big} {Data} ({Big} {Data})}}. \bibinfo{pages}{1483--1492}.
\newblock
\href{https://doi.org/10.1109/BigData.2016.7840756}{doi:\nolinkurl{10.1109/BigData.2016.7840756}}


\bibitem[Suter et~al\mbox{.}(2026)]%
        {suter2025terminology}
\bibfield{author}{\bibinfo{person}{Frédéric Suter}, \bibinfo{person}{Tainã Coleman}, \bibinfo{person}{İlkay Altintaş}, \bibinfo{person}{Rosa~M. Badia}, \bibinfo{person}{Bartosz Balis}, \bibinfo{person}{Kyle Chard}, \bibinfo{person}{Iacopo Colonnelli}, \bibinfo{person}{Ewa Deelman}, \bibinfo{person}{Paolo {Di Tommaso}}, \bibinfo{person}{Thomas Fahringer}, \bibinfo{person}{Carole Goble}, \bibinfo{person}{Shantenu Jha}, \bibinfo{person}{Daniel~S. Katz}, \bibinfo{person}{Johannes Köster}, \bibinfo{person}{Ulf Leser}, \bibinfo{person}{Kshitij Mehta}, \bibinfo{person}{Hilary Oliver}, \bibinfo{person}{J.-Luc Peterson}, \bibinfo{person}{Giovanni Pizzi}, \bibinfo{person}{Loïc Pottier}, \bibinfo{person}{Raül Sirvent}, \bibinfo{person}{Eric Suchyta}, \bibinfo{person}{Douglas Thain}, \bibinfo{person}{Sean~R. Wilkinson}, \bibinfo{person}{Justin~M. Wozniak}, {and} \bibinfo{person}{Rafael {Ferreira da Silva}}.} \bibinfo{year}{2026}\natexlab{}.
\newblock \showarticletitle{A terminology for scientific workflow systems}.
\newblock \bibinfo{journal}{\emph{Future Generation Computer Systems}}  \bibinfo{volume}{174} (\bibinfo{year}{2026}), \bibinfo{pages}{107974}.
\newblock
\showISSN{0167-739X}
\href{https://doi.org/10.1016/j.future.2025.107974}{doi:\nolinkurl{10.1016/j.future.2025.107974}}


\end{thebibliography}

\end{document}